\documentclass[useAMS,usenatbib]{mn2e}
\usepackage{graphicx,amssym,color}
\newcommand{\bc}{\begin{center}}
\newcommand{\ec}{\end{center}}

\title[Thick and Thin Disk] {
  Through Thick and Thin:  Kinematic and Chemical Components in the
  Solar Neighbourhood}
\author[Navarro et al]
{
Julio F. Navarro$^1$,
Mario G. Abadi$^2$, Kim A. Venn$^1$, K. C. Freeman$^3$, Borja Anguiano$^4$\\
$^1$Department of Physics and Astronomy, University of Victoria,
  Victoria BC, Canada\\
$^2$Observatorio Astron\'omico, Universidad Nacional de C\'ordoba,
C\'ordoba, Argentina\\
$^3$Research School of Astronomy and Astrophysics, The Australian National
University, Weston Creek ACT 2611, Australia\\
$^4$Astrophysikalisches Institut Potsdam, An der Sternwarte 16,
D-14482 Potsdam, Germany
}
\begin{document}

\date{Accepted 2010 ???? ??.
      Received 2010 ???? ??;
      in original form 2010 ???? ??}

\pagerange{\pageref{firstpage}--\pageref{lastpage}}
\pubyear{2010}

\maketitle

\label{firstpage}

\begin{abstract} We search for chemically-distinct stellar components
  in the solar neighbourhood using a compilation of published
  data. Extending earlier work, we show that when the abundances of
  Fe, $\alpha$ elements, and the $r$-process element Eu are considered
  together, stars separate neatly into two groups that delineate the
  traditional thin and thick disk components of the Milky Way.  The
  group akin to the thin disk is traced by stars with ${\rm
    [Fe/H]}>-0.7$ and [$\alpha$/Fe]$<0.2$. The thick disk-like group
  overlaps the thin disk in [Fe/H] but has higher abundances of
  $\alpha$ elements and Eu.  Stars in the range $-1.5<$[Fe/H]$<-0.7$
  with low [$\alpha/$Fe] ratios, however, seem to belong to a
  separate, dynamically-cold, non-rotating component that we associate
  with tidal debris. The kinematically-hot stellar halo dominates the
  sample for [Fe/H]$<-1.5$. These results suggest that it may be
  possible to define the main dynamical components of the solar
  neighbourhood using {\it only} their chemistry, an approach with a
  number of interesting consequences. With such definition, the
  kinematics of {\it thin disk} stars is found to be independent of
  metallicity: their average rotation speed remains roughly constant
  in the range $-0.7<{\rm [Fe/H]}<+0.4$, a result that argues against
  radial migration having played a substantial role in the evolution
  of the thin disk.  The velocity dispersion of the thin disk is also
  independent of [Fe/H], implying that the familiar increase in
  velocity dispersion with decreasing metallicity is the result of the
  increasing prevalence of the thick disk at lower metallicities,
  rather than of the sustained operation of a dynamical heating
  mechanism. The substantial overlap in [Fe/H] and, likely, stellar
  age, of the various components might affect other reported trends in
  the properties of stars in the solar neighbourhood. A purely
  chemical characterization of these components would enable us to
  scrutinize these trends critically in order to understand which
  result from accretion events and which result from secular changes
  in the properties of the Galaxy.
\end{abstract}


\section{Introduction}
\label{SecIntro}

The discovery that the vertical distribution of stars in the solar
cylinder is best approximated by two exponential laws suggested the
presence of two components of different scaleheights and ushered in
the concept of the Galactic thick disk \citep{Gilmore1983}. The
concept of separate ``thin'' and ``thick'' disk components in the
Milky Way, however, is useful insofar as they refer to different types
of stars; i.e., distinct in ways other than kinematics or spatial
distribution.  Since there is no {\it a priori} reason why the
vertical structure of galaxy disks cannot be more complex than a
single exponential, confirming the identity of the thick disk as a
component truly distinct from the thin disk requires additional
evidence.

In that regard, the finding that thick disk stars are substantially
older and more highly enriched in $\alpha$ elements than their thin
disk counterparts supports the two-component nature of the Galactic
disk
\citep{Gratton1996,Fuhrmann1998,Prochaska2000,Bensby2003,Gilmore1985,Reddy2006,Bensby2007,Fuhrmann2008}. The
thin and thick disks have metallicity distributions that overlap in
[Fe/H] but that differ, {\it at given} [Fe/H], in their kinematics,
age, and $\alpha$ content. This dichotomy in properties at fixed
metallicity requires adjustments to traditional models of chemical
evolution, which have invoked violent accretion events, as well as
episodic hiatus in star formation, to explain the data \citep[see,
e.g.,][]{Chiappini1997,Reddy2006}.  Such events may reset selectively
the heavy-element abundance of the ISM, disrupting the monotonic
trends with [Fe/H] expected in simple models of chemical evolution and
enabling better fits to the data.

The price to pay is one of increased model complexity. Did accretion
events supply mostly gas or stars to the disk? Were stars in the thick
disk brought into the Galaxy during the event or is the thick disk the
stirred-up remnant of an early thin disk? How many accretion events
took place? What were the masses of the accreted dwarfs? When did
these events take place?  Accretion models provide only rudimentary
answers to these key questions and, as a result, there is little
consensus in the community regarding the origin of the thick disk.

This state of affairs has been highlighted by a number of recent
papers, which argue that the chemo-dynamical evidence for accretion
events is weak and that all relevant data can be explained by
reconsidering the importance of radial migration of both gas and stars
during the evolution of the disk
\citep[][]{Haywood2008,Roskar2008,Schoenrich2009a}. Inspiration for
this work came from the realization by \citet{Sellwood2002} that
inhomogeneities in the disk do not just perturb disk stars
diffusively, easing them gradually into more and more eccentric
orbits, but can also transport, resonantly, stars across the disk
without increasing their eccentricity. Stars on nearly circular orbits
in the solar neighbourhood could therefore have formed at different
radii in the Galactic disk. The properties of local thin disk stars
may thus reflect large-scale gradients in the Galaxy rather than
different conditions in the solar neighbourhood at the time of their
formation.  Using a simple but plausible model that includes radial
migration, \citet{Schoenrich2009b} demonstrate that most available
data for the thin and thick disks can be accommodated without invoking
a violent accretion event at all.

Deciding between migration or accretion scenarios requires identifying
patterns in stellar properties that point to the presence of truly
distinct stellar component and assessing whether their properties are
the result of secular evolution or accretion events. In general,
secular evolution mechanisms such as radial migration should lead to
increased mixing, blurring the boundaries between components and
leaving dynamical and chemical imprints different from those predicted
by accretion scenarios.

Take, as an example, the wide range in metallicity spanned by stars on
nearly-circular orbits (the thin disk) in the solar neighbourhood. In
migration-based scenarios the metal-poor tail of the local thin disk
is populated by stars that formed further out in the Galaxy, while the
opposite holds for stars in the metal-rich tail. Stars at these two
extremes were born with very different angular momenta, and this
difference would be preserved in local samples even after their orbits
have migrated to the solar neighbourhood \citep{Schoenrich2009b}.
Therefore very metal-rich (poor) stars would be found in the solar
neighbourhood at relatively low (high) rotation velocities. A
relatively clean prediction of migration models is, then, the presence
of a negative correlation between mean rotation speed and metallicity
for stars in the thin disk.

Such correlations have been difficult to study because of the
widespread practice of assigning stars to the thick or thin disk
according to their kinematics. Although this may be an expeditious
procedure, it imposes an obvious bias that precludes searching for
correlations of the kind alluded to in the preceding paragraph. It
would be much more useful to devise a {\it purely chemical}
characterization of the various components in the Galaxy. After all,
chemistry is a relatively stable property of a star that relates it to the
conditions at the time/place of its birth, whereas its distance to the
center, vertical height, or spatial velocity are far less durable
features.

We explore these ideas here using a compilation of data for stars in
the solar neighbourhood with good estimates of their spatial motions
and reliable measurements of their heavy-element abundances. We focus on
spectroscopic samples where the abundance of individual elements, such
as Fe, Mg, Ti, and Ca, are also available. We use these data to apportion
stars to separate components on the basis of chemistry alone, and use
their kinematics to place constraints on their possible origins.

\section{Dataset}
\label{SecData}

We have used the compilation of \citet{Venn2004}, which includes
abundance data for solar neighbourhood stars with good spatial
motions, supplemented by data in the more recent papers of
\citet{Reddy2006}, \citet{Bensby2005} and \citet{Nissen2010}. This is
a heterogeneous compilation of data from many different sources.  We
have gone carefully through this compilation in order to remove
repeated stars and to bring all velocities to a consistent reference
frame, taking into account the fact that some papers use either right-
or left-handed UVW frames.  The sample spans a wide range in
metallicity ($-4<$[Fe/H]$<+0.5$), and there are reliable abundance
measurements of [Fe/H] and of the $\alpha$ elements Mg, Ti, and Ca,
for $743$ stars. Abundance measurements of the $r$-process element Eu
are also available for $306$ of those stars. We shall use these two
subsamples in what follows. As in \citet{Venn2004}, the spatial
velocities assume that the Sun moves in a Galactic reference frame
with (U,V,W)=($9$,$232$,$7$) km/s and that the LSR has $V_{\rm
  LSR}=220$ km/s.

\section{Results}
\label{SecRes}

\begin{figure}
\begin{center}
\includegraphics[width=1.0\linewidth,clip]{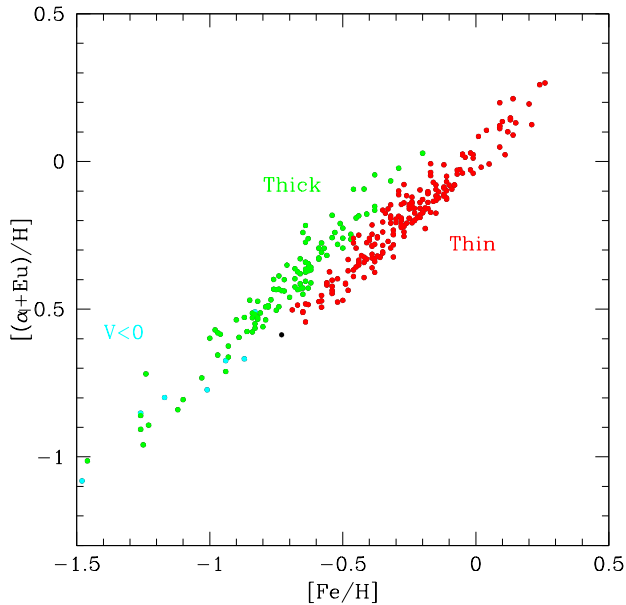}
\end{center}
\caption{Iron abundance vs an index constructed by averaging the
  abundance of the $\alpha$ elements (Mg, Ti, and Ca) plus the
  $r$-process element Eu. Note the presence of two well-defined
  ``families'' of stars each spanning a wide range in [Fe/H]. Stars in
  the family plotted in red (i.e., those with $[\alpha/{\rm Fe}]<0.2$,
  see Fig.~\ref{FigFeA}) are those traditionally assigned to the thin
  disk. Those in green are $\alpha$-enhanced stars ($[\alpha/{\rm
    Fe}]>0.2$) usually associated with the thick disk. Counterrotating
  stars unlikely to belong to either the thin or thick disks are shown
  in cyan.\label{FigFeAEu}}
\end{figure}

Fig.~\ref{FigFeAEu} shows, as a function of the iron content, [Fe/H],
an index measuring the average abundance of the $\alpha$ elements Mg,
Ti, and Ca, plus the $r$-process element Eu for all stars in our
sample.  The index is constructed by simply averaging the Mg, Ti, Ca,
and Eu abundances. As usual, all quantities are shown in log$_{10}$
units normalized to solar values.  Like the $\alpha$ elements, Eu is
thought to form mainly in massive stars, and it therefore may be
combined with the $\alpha$ elements to define an index that gauges
the importance of massive-star nucleosynthesis in a given stellar
component.

%%%%%%%%%%%%%%%%%%%%%%%%%%%%%%%%%%%%%%%%%%%
%%%%%%%%%%%%%%%%%%%%%%%%%%%%%%%%%%%%%%%%%%%
%%%%%%%%%%%%%%%%%%%%%%%%%%%%%%%%%%%%%%%%%%%
\begin{figure}
\begin{center}
\includegraphics[width=1.0\linewidth,clip]{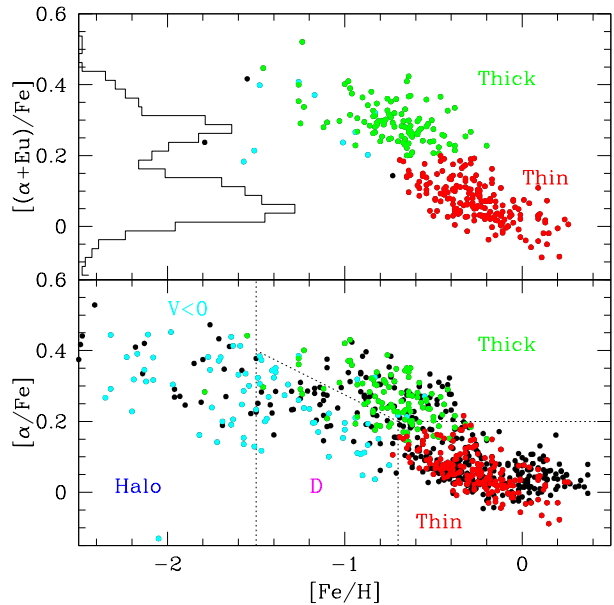}
\end{center}
\caption{{\it Top panel:} Same as Fig.~\ref{FigFeAEu}, but for the
  $\alpha+$Eu index normalized to Fe. The presence of the two families
  of stars hinted at in Fig.~\ref{FigFeAEu} is confirmed by the
  $[(\alpha+{\rm Eu)/Fe}]$ histogram.  We identify the bottom
  sequence, shown in red, with the thin disk, and the top sequence, in
  green, with the thick disk. The conditions used to define the thin
  disk are (i) ${\rm [Fe/H]}>-0.7$ and (ii) $[(\alpha+{\rm
    Eu)/Fe}]<0.2$. Stars in the region defined by [Fe/H]$>-1.5$
  and $[(\alpha+{\rm Eu)/Fe}]>0.2$ are those traditionally associated
  with the thick disk. Note that this condition allows in some
  counterrotating stars (${\rm V}<0$, shown in cyan) unlikely to be
  true members of the thick disk; the remainder are shown in
  black. {\it Bottom panel:} Same as top panel but for the index
  without Eu. Stars with available Eu are shown here using the same
  red and green colours as in the top panel. Stars in cyan are
  counterrotating ($V<0$) stars, unlikely to belong to either the thin
  or thick disks.  Note that counterrotating stars in the range
  $-1.5<$[Fe/H]$<-0.7$ have lower [$\alpha/$Fe] values than those of
  the thick disk. These stars appear to belong to a component (``D'')
  separate from the thin and thick disk or the traditional stellar
  halo, which we link to tidal debris. See text for further
  details.\label{FigFeA}}
\end{figure}
%%%%%%%%%%%%%%%%%%%%%%%%%%%%%%%%%%%%%%%%%%%

%%%%%%%%%%%%%%%%%%%%%%%%%%%%%%%%%%%%%%%%%%%
\begin{figure}
\begin{center}
\includegraphics[width=1.0\linewidth,clip]{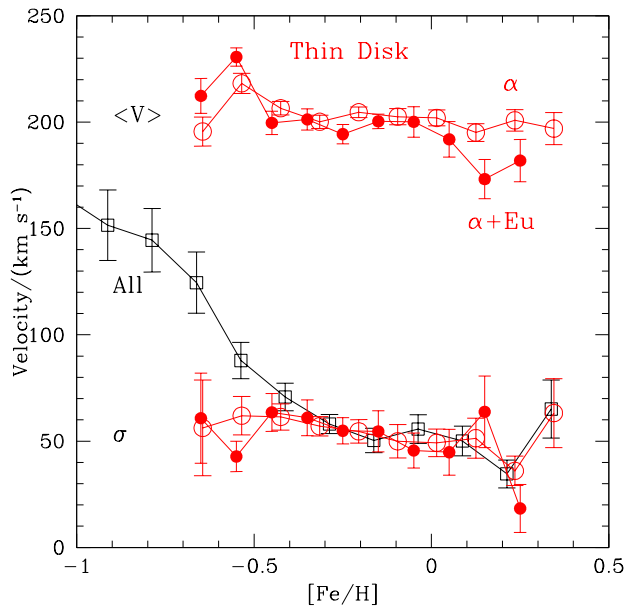}
\end{center}
\caption{Mean rotation velocity and velocity dispersion as a function
  of [Fe/H] for stars identified as belonging to the thin disk
  component in Fig.~\ref{FigFeA}. Filled circles correspond to stars
  with measured Eu abundances (top panel of Fig.~\ref{FigFeA}); open
  circles to all stars in the region labelled ``Thin'' in the bottom
  panel of Fig.~\ref{FigFeA}. Error bars indicate one-sigma bootstrap
  error estimates. Note that, although they span more than a decade in
  [Fe/H], {\it the average rotation velocity and velocity dispersion
    of thin disk stars is roughly independent of metallicity}. The
  open squares shows the $\sigma$-[Fe/H] correlation for all stars in
  our sample and illustrates the familiar increase of the total
  velocity dispersion ( $\sigma$) with
  decreasing [Fe/H].\label{FigFeVThin} }
\end{figure}
%%%%%%%%%%%%%%%%%%%%%%%%%%%%%%%%%%%%%%%%%%%
\subsection{A chemical definition of the thin disk}

Fig.~\ref{FigFeAEu} suggests that stars with measured Fe, $\alpha$ and Eu
abundances separate into two ``families'', one with high [Fe/H] and
low [($\alpha+$Eu)/Fe] (shown in red) and the other with lower
[Fe/H] and high [($\alpha+$Eu)/Fe] (green).  This separation is
confirmed in the top panel of Fig.~\ref{FigFeA}, which is identical to
Fig.~\ref{FigFeAEu} but for the index normalized to Fe. The
$[(\alpha+{\rm Eu})/{\rm Fe}]$ histogram, which should be relatively
free from selection biases, hints strongly at the presence of two
distinct components. This separation highlights the well-known $\alpha$
enhancement of thick disk stars relative to the thin disk, accentuated
by the addition of Eu to the $\alpha$ index. The data suggest that a
plausible boundary between the two families may be drawn at
[($\alpha+$Eu)/Fe]$=0.2$, the value that roughly marks the ``valley''
in the histogram.

Guided by this, we have chosen arbitrary but plausible boundaries to
define the two components in Fig.~\ref{FigFeAEu} and the top panel of
Fig.~\ref{FigFeA}, which we color in green and red, respectively. The
simple criteria: (i) [Fe/H]$>-0.7$, and (ii) [($\alpha$+Eu)/Fe]$<0.2$,
distinguishes well one of the two families of stars. This family
(coloured red) contains the Sun (which would be at the origin of the
plot) and contains mostly stars associated with the thin disk
as usually conceived. The minimum [Fe/H] boundary is suggested by the
fact that the most metal-rich counterrotating star in the sample
(certainly not a member of the thin disk) has [Fe/H]$\sim -0.7$.

The distinction between components blurs when including stars for
which Eu abundances are not available (see bottom panel of
Fig.~\ref{FigFeA}), but the sample more than doubles. In order to take
advantage of this larger sample, we {\it define the thin disk} by the
same criteria as above, namely
\begin{itemize}
\item (i) [Fe/H]$>-0.7$; and
\item (ii)  [$\alpha$/Fe]$<0.2$.
\end{itemize}
We emphasize that the criteria above are purely chemical. This differs
from the traditional practice of selecting thin disk stars by their
kinematics and might therefore include stars not expected in kinematic
definitions of the ``thin disk''. We shall hereafter call them ``thin
disk stars'', but this caveat should be kept in mind when comparing
our results with other work on the topic.

Both panels of Fig.~\ref{FigFeA} show the wide range in [Fe/H] spanned
by thin disk stars, as well as the strong correlation between the
abundance ratio and metallicity in this population; [$\alpha$/Fe]
and [($\alpha$+Eu)/Fe] decrease slightly but steadily with increasing
[Fe/H].

This trend suggests two possible interpretations. In standard chemical
evolution models it is reminiscent of a self-enriched population whose
star formation timescale is long compared with the lifetime of stars
that end their lives as supernovae type Ia (SNIa). These supernovae
return mostly Fe and little $\alpha$ to the ISM, and therefore the
ratio [$\alpha$/Fe] of successive generations of stars declines
steadily as [Fe/H] increases. 

In this interpretation, the formation of the metal-poor tail of the
thin disk precedes that of the metal-rich one since [Fe/H] is assumed
to march roughly monotonically with time.  One difficulty with this
interpretation is that the wide range in metallicity spanned by the
thin disk ($-0.7<$[Fe/H]$<+0.4$; see top panel of Fig.~\ref{FigFeA})
requires a protracted star formation history and implies a strong
correlation between age and metallicity.  This seems at odds with the
relatively weak age-[Fe/H] relation observed for thin disk stars.  One
must note here that stellar ages are notoriously difficult to
estimate, and that the strength of the predicted correlation depends
on assumptions about infall and star formation efficiency that are
difficult to pin down accurately. Despite these caveats, a number of
authors have concluded that the lack of a strong age-metallicity
relation signals the need to consider a different mechanism to explain
the properties of the thin disk \citep[see, e.g.,][and references
therein]{Haywood2008}.

These difficulties are circumvented in a second, different
interpretation of the properties of the local thin disk. As shown by
\citet{Schoenrich2009b}, radial migration can populate the vicinity of
the Sun with ``thin disk'' stars that formed either inside or outside
the Solar circle. The properties of local thin disk stars might not
trace the chemical history of the solar neighbourhood, but rather
reflect the spread in birth radii of such stars.  Given the
metallicity gradient in the Galaxy, this could explain naturally the
wide range in metallicity of thin-disk stars found today in the solar
neighbourhood.  The same modelling predicts that, once a steady state
is reached, additional enrichment at different radii would be balanced
by the infall of fresh metal-poor gas from the IGM. Because of higher
densities, star formation progresses nearer the center more intensely
relative to the amount of available gas, leading to lower
metallicities and higher [$\alpha/$Fe] ratios with increasing
Galactocentric birth radius. These gradients are today reflected
locally in the observed correlation between [Fe/H] and [$\alpha/$Fe].

One disadvantage of this scenario is that it requires a steeper
metallicity gradient in the ISM than is usually accepted
\citep[see][and references therein]{Maciel2007}, as well as careful
balancing of the timescales of star formation, radial migration, and
gas infall at different radii in order to explain the observed slope
of the [$\alpha/$Fe]-[Fe/H] correlation. Another potential challenge
may arise from the fact that at least some metal-rich stars in the
inner Galaxy seem to be $\alpha$-enriched \citep{Melendez2008} whereas
those in the solar neighbourhood are not.

One potential test of the migration scenario was discussed in the
Introduction, and concerns the prediction of a negative correlation
between metallicity and rotation speed {\it in the thin disk}. Reading
off from Fig. 4 in \citet[][]{Schoenrich2009b}, the mean rotation
speed is expected to change from roughly $250$ to $160$ km/s as the
metallicity in the thin disk increases from [Fe/H]$=-0.7$ to $+0.5$.

As shown in Fig.~\ref{FigFeVThin}, this prediction finds some support
in the data.  Filled circles here refer to the $\alpha+$Eu data shown
in the top panel of Fig.~\ref{FigFeA} whereas open symbols correspond
to the much larger sample of stars without Eu data (bottom panel of
Fig.~\ref{FigFeA}).  The mean rotation speed of the thin disk (defined
by the above conditions) decreases somewhat with [Fe/H] in the
$\alpha+$Eu dataset but the trend is weaker than expected from
\citet[][]{Schoenrich2009b} and the error bars are large.

Furthermore, when the larger sample without Eu is considered (open
circles) the mean rotation speed is actually {\it independent} of
metallicity.  It appears as if the mean rotation speed, as well as the
velocity dispersion, remain roughly constant over the whole range in
[Fe/H] spanned by the thin disk.  The constant velocity dispersion of
the thin disk should be contrasted with the familiar increase in
$\sigma$ with decreasing [Fe/H] \citep{Stroemgren1987}; this is shown
by the open squares in Fig.~\ref{FigFeVThin} which corresponds to {\it
  all} stars in our compilation.

The kinematic invariance of the thin disk with metallicity is a
somewhat surprising result for either of the two scenarios advanced
above. The absence of a $\langle$V$\rangle$-[Fe/H] correlation
disfavours the migration scenario; on the other hand, {\it in-situ}
self-enrichment scenarios would predict the velocity dispersion to
rise with decreasing [Fe/H] since metal-poor stars would be, on
average, older, and therefore exposed for longer to gravitational
heating by inhomogeneities in the Galactic potential. The data in
Fig.~\ref{FigFeVThin} disfavours such gradual heating; if present, the
heating mechanism must operate promptly and saturate quickly, as has
been suggested in the past
\citep{Stroemgren1987,Freeman1991,Quillen2000,Soubiran2008}.
Inefficient heating would actually be easier to reconcile with
theoretical models, which have struggled to explain the rapid increase
in velocity dispersion with age inferred from earlier observations
\citep[see, e.g.,][]{Wielen1977,Lacey1984,Jenkins1990,Aumer2009}.

%%%%%%%%%%%%%%%%%%%%%%%%%%%%%%%%%%%%%%%%%%%
\begin{figure}
\begin{center}
\includegraphics[width=1.0\linewidth,clip]{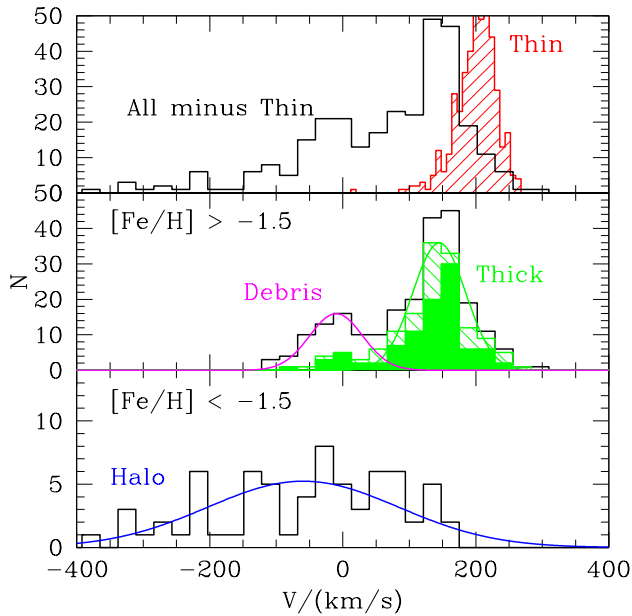}
\end{center}
\caption{{\it Top panel:} The open histogram shows the distribution of
  the rotation velocity (V component) for all stars in the
  sample. (Units are number of stars per bin.) The shaded histogram
  corresponds to those in the thin disk, as defined in the bottom
  panel of Fig.~\ref{FigFeA}. The remainder are shown by the open
  histogram. {\it Bottom panel:} V-distribution of stars in the
  metal-poor tail, i.e., [Fe/H]$<-1.5$.  This illustrates the velocity
  distribution of the slowly-rotating, dynamically-hot stellar
  halo. The blue curve illustrates the best fitting gaussian, with
  $\langle V \rangle=-60$ km/s and $\sigma_V=144$ km/s. {\it Middle
    panel:} V-distribution of stars with [Fe/H]$>-1.5$ but {\it
    excluding} the thin disk. The distribution is strongly
  non-gaussian, and shows two well-defined peaks; one at $V\sim 0$
  km/s and another at $V\sim 160$ km/s. The latter is well traced by
  stars in the ``thick disk'' component identified in the top panel of
  Fig.~\ref{FigFeA} (with Eu; filled green histogram).  Including
  stars without Eu measurements but of comparable [$\alpha/$Fe] ratios
  (the ``Thick'' region in the bottom panel of Fig.~\ref{FigFeA})
  results in the shaded green histogram.  Note that stars near the
  $V\sim 0$ peak are almost exclusively those in the debris (``D'')
  region of Fig.~\ref{FigFeA}. They define a non-rotating,
  dynamically-cold component distinct from either the thick disk and
  the stellar halo.  Both the debris and thick disk components are
  relatively cold dynamically; the V-distribution can be well
  approximated by the sum of two gaussians with similar velocity
  dispersion, of order $\sim 40$ km/s (see green and magenta solid
  curves).
  \label{FigVhisto} }
\end{figure}
%%%%%%%%%%%%%%%%%%%%%%%%%%%%%%%%%%%%%%%%%%%

\subsection{A chemical definition of the thick disk}

If our interpretation is correct, then the increase in velocity
dispersion with decreasing metallicity for stars in the vicinity of
the Sun must result from the increased prevalence of the thick disk at
low metallicity. Indeed, as mentioned in Sec.~\ref{SecIntro}, it is
generally agreed that the thick disk is metal-poor; lags the thin disk
in rotation speed; and has a higher velocity dispersion. This is shown
in the top panel of Fig.~\ref{FigVhisto}, where we compare the
distribution of the rotation speed (V component) of all stars in our
sample with that of the thin disk.

The V distribution of stars {\it not} in the thin disk is complex, and
hints at the presence of distinct dynamical components. It shows two
well defined peaks, one at V$\sim 160$ km/s and another at $V\sim 0$
km/s, as well as a tail of fast counterrotating stars at
highly-negative values of V. The first peak corresponds to a
rotationally-supported structure: the traditional thick disk. Stars
belonging to this rotating component seem to disappear from our sample
when only metal-poor stars with ${\rm [Fe/H]} < -1.5$ are considered
(bottom panel of Fig.~\ref{FigVhisto}). The latter trace the
classical, kinematically-hot, metal-poor stellar halo: the V distribution is
consistent with a gaussian with velocity dispersion $\sigma_V\sim 144$
km/s (shown in the bottom panel with a blue curve).

Interestingly, the V distribution of non-thin-disk stars with ${\rm
  [Fe/H]}>-1.5$ (middle panel of Fig.~\ref{FigVhisto}) shows even more
clearly the double-peak structure noted above. The peak at ${\rm V}
\sim 160$ km/s is well traced by the stars identified with the thick
disk in the $\alpha+$Eu panel of Fig.~\ref{FigFeA}, shown by the
solid-shaded histogram in the middle panel of Fig.~\ref{FigVhisto}.

Indeed, the peak is traced almost exclusively by stars with high
values of [$\alpha$/Fe]. This is shown by the shaded green histogram,
which corresponds to all stars in the region labelled ``Thick'' in the
bottom panel of Fig.~\ref{FigFeA}. The V distribution of these stars
is well approximated by a gaussian with $\langle$V$\rangle=145$ km/s
and $\sigma_V=40$ km/s that accounts for nearly all stars with $V>100$
km/s.

The association between the thick disk and ``high-$\alpha$'' stars is
reinforced by inspecting the location of counterrotating stars in
Fig.~\ref{FigFeA} (shown in cyan). These stars, which clearly do not
belong to a rotationally-supported structure like the thick (or thin)
disk, are evenly distributed among stars with ${\rm [Fe/H]}<-1.5$ but
shun the ``high-$\alpha$'' region in the range $-1.5<{\rm
  [Fe/H]}<-0.7$. Of the $38$ counterrotating stars in our sample with
$-1.5<{\rm [Fe/H]}<-0.7$, only $3$ lie above the dotted line that
delineates the ``Thick'' region in the bottom panel of Fig~\ref{FigFeA}.

 It is tempting therefore to adopt a
purely chemical definition of the thick disk in terms of ${\rm [Fe/H]}$ and
$[\alpha/{\rm Fe}]$: 
\begin{itemize}
\item (i) ${\rm [Fe/H]}>-1.5$;

\item (ii) $[\alpha/{\rm Fe}]>0.2-({\rm [Fe/H]}+0.7)/4$ for
$-1.5<[{\rm Fe/H}]<-0.7$;

\item (iii) $[\alpha/{\rm Fe}]>0.2$ for ${\rm [Fe/H]}>-0.7$.
\end{itemize}

If ou analysis is correct, then the thick disk emerges as a
chemically and kinematically coherent component that spans a wide
range in metallicity ($-1.5<$[Fe/H]$<-0.3$) and contains mainly stars
highly enriched in $\alpha$ elements. Stars in this component have
($\langle$U$\rangle$,$\langle$V$\rangle$,$\langle$W$\rangle$)$=(10,133,-2)$
km/s, and ($\sigma_U$,$\sigma_V$,$\sigma_W$)$=(95,61,61)$ km/s.

The average V lags below the $\sim 160$ km/s peak because the chemical
definition allows in a few stars with discrepant velocities (some with
negative V) more closely associated with the ${\rm V}\sim 0$ peak than
with a rotationally supported structure like a thick disk. This
suggests that the chemical definition of the thick disk proposed above
is not perfect, and that it includes spurious stars belonging to a
different, non-rotating component. We turn our attention to that
component next.

\subsection{Tidal debris in the stellar halo}

The last feature of note in the top and middle panels of
Fig.~\ref{FigVhisto} is the peak centered at V $\sim 0$. These are
stars with no net sense of rotation around the Galaxy and with a
surprisingly low velocity dispersion. (The gaussian fit to the low-V
tail shown in magenta in the middle panel of Fig.~\ref{FigVhisto} has
a dispersion of just $40$ km/s.) Their vertical (W) velocity
distribution (computed for stars with ${\rm V}<50$ km/s to better
isolate the peak\footnote{Lifting this restriction has only a small
  impact on the dispersion; $\sigma_W\sim 90$ km/s for {\it all} stars in
  the ``D'' region of Fig.~\ref{FigFeA}.}) is also indistinguishable from that of the thick
disk component discussed above, with $\langle$W$\rangle\sim 0$ km/s
and $\sigma_W\sim 72$ km/s.  On the other hand, their U velocity
dispersion is large ($\sim 180$ km/s). Further, the U distribution
(not shown) is rather flat, with hints of two, symmetric peaks, one at
${\rm U}\sim -250$ km/s and another at ${\rm U} \sim +250$ km/s.

As discussed by \citet{Meza2005}, this is the kinematic structure
expected for a tidal stream originating in a dwarf galaxy whose
orbital plane at the time of disruption was coincident with the plane
of the Galaxy. The low V and W velocity dispersions are easily
explained in this scenario, since such stream would be, locally,
kinematically cold and confined to the orbital plane. The
double-peaked U distribution also arises naturally in this scenario if
stream stars have apocentric radii outside the solar circle and
pericentric radii inside the solar circle. Stream stars in the solar
neighbourhood are therefore either going to their apocenter with
large, positive U or coming from their apocenter, with symmetric
$-{\rm U}$.

\begin{figure}
\begin{center}
\includegraphics[width=1.0\linewidth,clip]{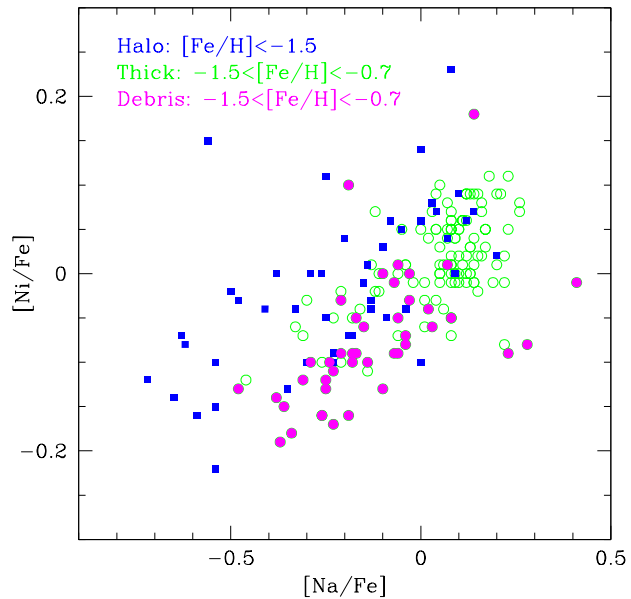}
\end{center}
\caption{[Na/Fe] vs [Ni/Fe] correlation for all stars in our sample
  with [Fe/H]$<-0.7$. Blue squares correspond to [Fe/H]$<-1.5$ ``classical halo''
  stars; green open circles to $\alpha$-rich ``thick disk'' stars with
  $-1.5<{\rm [Fe/H]}<-0.7$, and magenta filled circles to stars in the
  ``debris'' (``D'') region of Fig.~\ref{FigFeA}. Note how these three
  components separate neatly in the Na vs Ni   plane and, in
  particular, the tight scatter around the mean trend of ``debris''
  stars. This strengthens our conclusion that they correspond to three  components of distinct origin.
  \label{FigNaNi} }
\end{figure}

This suggests that most non-thin-disk stars with [Fe/H]$>-1.5$ and
low, but still enhanced relative to solar, [$\alpha$/Fe] (previously
thought to belong to the classical halo) belong to this new
component. Guided by \citet{Nissen2010}, we inspect the Na and Ni
content of such stars for supporting evidence of this conclusion. This
is shown in Fig.~\ref{FigNaNi}, where we show [Na/Fe] vs [Ni/Fe] for
all stars in our sample with ${\rm [Fe/H]}<-0.7$. Blue squares
correspond to the very metal poor stars in our sample (${\rm
  [Fe/H]}<-1.5$); green open circles are stars in the $\alpha$-rich
``Thick'' region of Fig.~\ref{FigFeA}, and magenta filled circles
denote stars in the $\alpha$-poor ``debris'' (``D'') region of
Fig.~\ref{FigFeA}. The three groups separate clearly in the Na-Ni
plane, supporting our claim that the ``debris'' component is truly
distinct from the thick disk and from the metal-poor ``classical''
halo.

Using the same ${\rm [Fe/H]}$ and $[\alpha/{\rm Fe}]$ parameters as above, we can characterize
``debris'' stars in the [$\alpha$/Fe] vs [Fe/H] plane (region labelled
``D'' in Fig.~\ref{FigFeA}) by
\begin{itemize}
\item (i) $-1.5<{\rm [Fe/H]}<-0.7$; and

\item (ii) $[\alpha/{\rm Fe}]<0.2-({\rm [Fe/H]}+0.7)/4$.
\end{itemize}
Our conclusion agrees with that of \citet{Nissen2010}, who studied a
large spectroscopic sample of ``halo'' stars and argued, in agreement
with our analysis, that most metal-rich ``$\alpha$-poor'' halo stars
are indeed tidal debris from disrupted dwarfs. Our debris population
is reminiscent of the population identified by \citet{Morrison2009}.

%%%%%%%%%%%%%%%%%%%%%%%%%%%%%%%%%%%%%%%%%%%
%%%%%%%%%%%%%%%%%%%%%%%%%%%%%%%%%%%%%%%%%%%

\section{Summary and Discussion}
\label{SecDisc}

The preceding analysis suggests that apportioning the various
components of the Galaxy according to purely chemical criteria is both
possible and fruitful. The definition of the thin disk in the
[($\alpha+$Eu)/Fe] vs [Fe/H] plane is particularly straightforward,
and suggests that the kinematics of the thin disk is invariant with
metallicity. This is an intriguing result  unexpected
in migration-based scenarios for the chemo-dynamical
evolution of the thin disk.  It implies that the familiar increase in
velocity dispersion with decreasing metallicity
\citep[][]{Stroemgren1987} is the result of the increased prevalence
of the thick disk at lower metallicities, rather than of the sustained
operation of a dynamical heating mechanism. If confirmed, the
kinematic invariance of the thin disk with metallicity will place
strong constraints on the formation of the Galactic disk and on the
role of accretion events, in situ formation, and/or migration.

The ``thick disk'' can also be charted in the [$\alpha$/Fe] vs
[Fe/H] plane. As reported in earlier work, it seems to contain mainly
stars highly enriched in $\alpha$ elements. It shows as a separate
dynamical component in rotation speed, with the bulk of its stars rotating
at ${\rm V} \sim 160$ km/s. A simple criterion in $\alpha$ content
isolates most of these stars, although a few outliers with nearly
zero, or negative, V velocities are also included. The latter might
very well be contaminants from a different population that a crude
boundary in the  $\alpha$-Fe plane  is unable to weed out. 

The inclusion of additional heavy elements in the defining criteria
might enable a cleaner characterization of the thick disk. If that
were possible, questions such as whether the thick disk shows evidence
of self-enrichment, or whether correlations between metal content and
velocity dispersion are present, could be addressed. This would allow
us to distinguish between migration and accretion models and, among
the latter, between those where the bulk of thick disk stars were
either accreted or simply stirred.

A substantial fraction of stars in the range $-1.5<{\rm [Fe/H]}<-0.7$
seem to belong to a dynamically-cold, non-rotating component with
properties consistent with those of a tidal stream. These are mainly
stars of low-$\alpha$ content, comparable in that regard to individual
stars in many of the satellite companions of the Milky Way \citep[see,
e.g.,][]{Venn2004,Tolstoy2009}: the low-[Fe/H], $\alpha$-poor region
should be a good hunting ground for the remnants of accretion events.

The V-distribution associated with the stream or ``debris'' seems to
peak at slightly negative V ($\sim -10$ km/s). This implies that
most stars in the stream have angular momenta similar to that of the
globular cluster $\omega$Cen \citep{Dinescu1999}. It is therefore
tempting to associate these stars with the parent galaxy of this
massive cluster, long suspected to be the survivor of the disruption
of a dwarf galaxy in the Galactic potential
\citep{Freeman1993}. Supporting evidence comes from $\omega$Cen's low
vertical velocity, as well as from the overlap in metallicity between
$\omega$Cen and stars in the putative stream \citep[see][for a full
discussion]{Meza2005,Nissen2010}.

Although the orbital parameters of the stream might suggest a link
with the globular cluster $\omega$Cen, we caution that evidence of a
true relation between those stars and $\omega$Cen is circumstantial at
best. For example, the metallicity distribution in $\omega$Cen peaks
at ${\rm [Fe/H]}=-1.7$ \citep{Smith2004} and have distinct elemental
abundances \citep{Norris1995,Johnson2010}, so it might be useful to seek evidence
for a stream in stars of similar metallicity and peculiar abundance ratios when
larger samples become available \citep[see, e.g.,][for some progress
in this direction]{WyliedeBoer2010}.

One corollary of this finding is that very few stars with ${\rm
  [Fe/H]}>-1.5$ in our sample seem to belong to the classical,
dynamically-hot stellar halo. Recent work has suggested that halo
stars in this ``metal-rich'' tail might actually belong to a distinct
``inner halo'' component \citep[see, e.g.,][and references
therein]{Carollo2007}.  The connection between this and
the stream we advocate above is, however, unclear, not least because
the ``inner halo'' component is reported to co-rotate with the Sun,
whereas our putative stream counterrotates slowly around the Galaxy.

The``inner halo'' identification typically relies mainly on estimates
of [Fe/H] and lacks information on the abundance of individual
elements. On the other hand, our sample is relatively small in
comparison and has potentially a number of selection biases. Thus the
possibility that our stream constitutes a small subset of the inner
halo remains. 

It might be possible to test these ideas by examining other abundance
ratios, such as the neutron-capture elements [Ba/Y] or [Ba/Eu]. This
is because these elements are contributed in different amounts by AGB
stars that undergo slow neutron capture nucleosynthesis during the
thermal pulsing stages and by massive stars that undergo rapid neutron
capture nucleosynthesis during supernova explosion. Indeed, variations
in those ratios have already been seen in individual stars of dwarf
galaxies \citep[see Figure 14 in][]{Tolstoy2009}.

Overall, our success in dividing and assigning stars of the solar
neighbourhood to families of distinct chemistry and kinematics seems
to favour models where accretion events have played a significant role
in the formation of the Galaxy rather than models, such as those based
on migration, where secular evolutionary mechanisms rule. We hasten to
add, however, that the criteria to separate components proposed here
are imperfect, and that our conclusions are based on small and
heterogeneous samples. These samples likely conceal a number of biases
which can only be revealed and lifted by a concerted effort to survey
a large, volume-limited, kinematically-unbiased sample of stars with
the high-resolution spectra needed to measure the abundance of
individual heavy elements. The planned {\tt HERMES}
survey\footnote{\tt www.aao.gov.au/AAO/HERMES} should be a major first
step in this regard. This undoubtedly ambitious endeavour would allow
us to dissect chemically and kinematically the solar neighbourhood and
to learn the true provenance of the many stellar families that today
call this small place of the Galaxy home.

%%%%%%%%%%%%%%%%%%%%%%%%%%%%%%%%%%%%%%%%%%%%%%%%%%%%%%%%%%%%%%%%%%%
\section*{Acknowledgements}

\bsp
\label{lastpage}

\bibliographystyle{mn2e}
\bibliography{master}

\end{document}